\documentclass[aps,twocolumn,superscriptaddress, floatfix,prl]{revtex4}
\usepackage{graphicx,amsmath}
\usepackage{graphicx}
\usepackage{dcolumn}
\usepackage{bm}
\usepackage{url}
\usepackage{color}
\usepackage{floatflt}
\usepackage{hyperref}
\usepackage{amssymb,amsmath}
\usepackage{graphicx}
\usepackage{graphics}
\usepackage{wrapfig}
\usepackage{subfigure}
\usepackage{float}

\def\bu{\boldsymbol{u}}

\begin{document}

\title{Mushroom spore dispersal by convectively-driven winds}
\author{E. Dressaire}
\affiliation{New York University Polytechnic School of Engineering, Brooklyn, NY 11201}
\author{Lisa Yamada}
\affiliation{Dept. of Engineering, Trinity College, CT 06106} 
\author{Boya Song}
\affiliation{Dept. of Mathematics, UCLA, Los Angeles, CA 90095, USA} 
\author{Marcus Roper}
\affiliation{Dept. of Mathematics, UCLA, Los Angeles, CA 90095, USA} 
\affiliation{Dept. of Biomathematics, UCLA, Los Angeles, CA 90095, USA}

\begin{abstract}
Thousands of fungal species rely on mushroom spores to spread across landscapes. It has long been thought that spores depend on favorable airflows for dispersal -- that active control of spore dispersal by the parent fungus is limited to an impulse delivered to the spores to carry them clear of the gill surface. Here we show that evaporative cooling of the air surrounding the mushroom pileus creates convective airflows capable of carrying spores at speeds of centimeters per second. Convective cells can transport spores from gaps that may be only a centimeter high, and lift spores ten centimeters or more into the air. The work reveals how mushrooms tolerate and even benefit from crowding, and provides a new explanation for their high water needs.
\end{abstract}
\maketitle

\section{Introduction}
Rooted in a host organism or patch of habitat such as a dead log, tens of thousands of species of filamentous fungi rely on spores shed from mushrooms and passively carried by the wind to disperse to new hosts or habitat patches. A single mushroom is capable of releasing over a billion spores per day \cite{kadowaki2010periodicity}, but it is thought that the probability of any single spore establishing a new individual is very small \cite{nagarajan1990long,Galante_11}. Nevertheless in the sister phylum of the mushroom-forming fungi, the Ascomycota, fungi face similarly low likelihoods of dispersing successfully, but spore ejection apparatuses are highly optimized to maximize spore range \cite{Roper_10,roper2008explosively,Fritz_13}, suggestive of strong selection for adaptations that increase the potential for spore dispersal.

Spores disperse from mushrooms in two phases \cite{money1998more}: a powered phase, in which an initial impulse delivered to the spore by a surface tension catapult carries it clear of the gill or pore surface, followed by a passive phase in which the spore drops below the pileus and is carried away by whatever winds are present in the surrounding environment. The powered phase requires feats of engineering both in the mechanism of ejection \cite{Pringle_05, Noblin_09,stolze2009adaptation} and in the spacing and orientation of the gills or pores \cite{Buller_book,Fischer_10_2}. However, spore size is the only attribute whose influence on the passive phase of dispersal has been studied \cite{Norros_14}. Spores are typically less than 10$\mu$m in size, so can be borne aloft by an upward wind of only 1 cm/s \cite{Buller_book}. Buller claimed that such wind speeds are usually attained beneath fruiting bodies in Nature \cite{Buller_book}: Indeed peak upward wind velocities under grass canopies are of order 0.1-1 cm/s \cite{Aylor_90}. However even if the peak wind velocity in the mushroom environment is large enough to lift spores aloft: 1. the average vertical wind velocity is zero, with intervals of downward as well as upward flow and 2. mushrooms frequently grow in obstructed environments, such as close to the ground or with pilei crowded close together (Fig 1 A-B). The pileus traps a thin boundary layer of nearly still air, with typical thickness $\delta\sim \sqrt{\frac{\nu L}{U}}$ \cite{Batchelorbook}, where $U$ is the horizontal wind velocity, $L$ the size of the pileus, and $\nu$ the viscosity of air \cite{Batchelorbook}; no external airflow can penetrate into gaps narrower than $2\delta$. For typically sized mushrooms under a grass canopy (with $L=10$ cm, $U=1-10$ cm/s, $\nu=1.5\times 10^{-5}$ m$^2$s$^{-1}$), we find that $2\delta\approx 6-20$ mm. If the gap thickness between pileus and ground is smaller than $2\delta$ then no external wind will penetrate into the gap.

\section{Results and Discussion}\label{sec:exp_results}

{\bf Spores can disperse from thin gaps beneath pilei without external winds.} We analyzed spore deposition beneath cultured mushroom (shiitake; {\it Lentinula edodes}, and oyster; {\it Pleurotus ostreatus}, sourced from CCD Mushroom, Fallbrook, CA) as well as wild-collected {\it Agaricus californicus}. Pilei were placed on supports to create controllable gap heights beneath the mushroom, and placed within boxes to isolate them from external airflows. We measured spore dispersal patterns by allowing spores to fall onto sheets of transparency film, and photographing the spore deposit (Fig. 1C). In all experiments, spore deposits extended far beyond the gap beneath the pileus. Spores were deposited in asymmetric patterns, both for cultured and wild-collected mushrooms (Fig. 1D). Using a laser light sheet and high speed camera (see Materials and Methods), we directly visualized the flow of spores leaving the narrow gap beneath a single mushroom (Fig. 1E and Movie S1): our videos show that spores continuously flow out from thin gaps, even in the absence of external winds. 

What drives the flow of spores from beneath the pileus? Some ascomycete fungi and ferns create dispersive winds by direct transfer of momentum from the fruiting body to the surrounding air. For example, some ascomycete fungi release all of their spores in a single puff; the momentum of the spores passing through the air sets the air into motion \cite{Roper_10}. Fern sporangia form over-pressured capsules that rupture to create jets of air \cite{Whitaker_10}. However, the flux of spores from a basidiomycete pileus is thousands of times smaller than for synchronized ejection by a ascomycete fungus, and pilei have no known mechanism for storing or releasing pressurized air. The only mechanism that we are aware of for creating airflows without momentum transfer is by the manipulation of buoyancy -- an effect that underpins many geophysical flows \cite{pedlosky1982geophysical} and has recently been tapped to create novel locomotory strategies \cite{mercier2014self}.

{\bf Mushrooms evaporatively cool the surrounding air.} Many mushrooms are both cold and wet to the touch \cite{Arora_86};  the expanding soft tissues of the fruit body are hydraulically inflated, but also lose water quickly (Fig. 2A). 
We made a comparative measurement of water loss rates from living mushrooms and plants. The rate of water loss from mushrooms greatly exceeds water loss rates for plants, which use stomata and cuticles to limit evaporation (Fig. 2B). For both plants and pilei, rates of evaporation were larger when tissues were able to actively take in water via their root-system / mycelium, than for cut leaves or pilei. However, the pilei lost water more quickly than all species of plants surveyed under both experimental conditions. In fact, evaporation rates from cut mushrooms were comparable to a sample of water agar hydrogel (1.5\% wt/vol agar), while evaporation rates from mushrooms with intact mycelia, were twofold larger (Fig 2B). Taken together, these data suggest that pilei are not adapted to conserve water as effectively as the plant species analyzed. 

The high rates of evaporation lead to cooling of the air near the mushroom, and may have adaptive advantage to the fungus. Previous observations have shown that evaporation cools both the pileus itself and the surrounding air by several degrees Celsius \cite{Husher_99}. Specifically, since latent heat is required for the change of phase from liquid to vapor, heat must continually be transferred to mushroom from the surrounding air. We compared the ambient temperature of the air between 20 cm and 1 m away from {\it P. ostreatus} pilei with intact mycelia, with temperatures in the narrow gaps between and beneath pilei, using a Traceable liquid/gas probe (Control Company, Forestwood, TX).  We found that gap temperatures were consistently 1-2$^\circ$C cooler than ambient (Fig. 2C). The surface temperature of the pileus, measured with a Dermatemp infra-red thermometer (Exergen, Watertown, MA) was up to 4$^\circ$C cooler than ambient, consistent with previous observations (\cite{Husher_99} and Fig. 2C). Evaporation alone can account for these temperature differences: in a typical experimental run, a pileus loses water at a rate of  $3\times 10^{-5}$ kg/m$^2$s (comparable with the data of \cite{Mahajan_08}). Evaporating this quantity of water requires $E_{\rm vap} \approx 70$ W/m$^2$ of vaporization enthalpy. At steady state the heat flux to the mushroom must equal the enthalpy of vaporization; Newton's law of cooling gives that the heat flux (energy/area) will be proportional to the temperature difference, $\Delta$ between the surface of the mushroom and the ambient air: $E_{\rm vap} = h\Delta T$ where $h=10-30$ W/m$^2$\,$^o$C is a heat transfer coefficient \cite{baehr2011heat}. From this formula we predict that $\Delta T \approx 2.5-7^\circ$C, in line with our observations. 

{\bf Increasing air density by cooling produces dispersive currents}. Cooling air from the laboratory ambient ($T=18^\circ$C) down to $T=16^\circ$C increases the density of the air by $\Delta \rho_T=\alpha\Delta T = 0.008$ kg\,m$^{-3}$, where $\alpha$ is the coefficient of expansion of air. Cold dense air will tend to spread as a gravity current \cite{Huppert_06}, and an order of magnitude estimate for the spreading velocity of this gravity current from a gap of height $h=1$ cm is given by von Karman's law: $U_g \sim \sqrt{\frac{2 \Delta \rho_T g h}{\rho_0}} \approx 4$cm/s. Although the air beneath the pileus is laden with spores, spore weight contributes negligibly to the creation of dispersive winds: in typical experiments, spores were released from the pileus at a rate of $q=$540$\pm$ 490 spores/cm$^2$ s (data from $24$ {\it P. ostreatus} mushrooms). If the mass of a single spore is $m_s=5\times 10^{-13}$kg and its sedimentation speed is $v_s = 10^{-3}$ m/s (see discussion preceding Equation (1)), then the contribution of spores to the density of air beneath the pileus is $\Delta \rho_S = m_s q/v_s = 0.003$ kg\,m$^{-3}$, less than half of the density increase produced by cooling $\Delta \rho_T$. Indeed, water evaporation, rather than spores and the water droplets that propel them \cite{Buller_book}, constitute most of the mass lost by a mushroom. To prevent spore ejection, we applied a thin layer of petroleum jelly to the gill surfaces of cut mushrooms. Treated mushrooms lost mass at a statistically indistinguishable rate to cut mushrooms that were allowed to shed spores (Fig. 2B).

Although our experiments were performed in closed containers to exclude external airflows it is still possible that spore deposit patterns were the result of convective currents created by temperature gradients in the lab, rather than airflows created by the mushroom itself. To confirm that spores were truly dispersed by airflows created by the mushroom we rotated the mushroom either 90$^\circ$ or 180$^\circ$ halfway through the experiment and replaced the transparency sheet. Since the box remained in the same orientation and position in the lab, we would expect that if spores are dispersed by external airflows, the dispersal pattern would remain the same relative to the lab. In fact we found consistently that the direction of the dispersal current rotated along with the mushroom (see Figure S1), indicating that mushroom generated air-flows are dispersing spores.

We explored how the distance dispersed by spores depended on factors under the control of the parent fungus. The distance spores dispersed from the pileus increased in proportion to the square of the thickness of the gap beneath the pileus ($R^2 = 0.90$, Fig. 3). However, we found no correlation between spore dispersal distance and the diameter of the pileus or the rate at which spores were produced ($R^2=0.07$ and $R^2=0.17$ respectively, Figure S2).

Spores were typically deposited around mushrooms in asymmetric patterns, suggesting that one or two tongues of spore laden air emerge from under the pileus, and spores do not disperse symmetrically in all directions (Fig. 1C). These tongues of deposition were seen in wild-collected as well as cultured mushrooms (Fig. 1D). We dissected the dynamics of one of these tongues by building a two dimensional simulation of the coupled temperature and flow fields around the pileus. Although real dispersal patterns are three dimensional, these simulations approximate the 2D dynamics along the symmetry plane of a spreading tongue. In our simulations we used a Boussinesq approximation for the equations of fluid motion and model spores as passive tracers, since their mass contributes negligibly to the density of the gravity current. Initially we modeled the pileus by a perfect half-ellipse whose diameter (4 cm) and height (0.8 cm) matched the dimensions of a {\it L. edodes} pileus used in our experiments. However, if cooling was applied uniformly over the pileus surface then spores dispersed weakly (Fig. 4A). Weak symmetric dispersal can be explained by conservation of mass: cold outward flow of spore-laden air must be continually replenished with fresh air drawn in from outside of the gap. In a symmetric pileus, the cool air spreads along the ground and inflowing air travels along the under-surface of the pileus. So initially on leaving the gills of the mushroom, spores are drawn inward with the layer of inflowing warm air; and only after spores have sedimented through this layer into the cold outflow beneath it do they start to travel outward (Fig 4A, upper panel).  

{\bf Asymmetric airflows are necessary for dispersal.} To understand how mushrooms can overcome the constraints associated with needing to maintain both inflow and outflow, we performed a scaling analysis of our experimental and numerical data. Although the buoyancy force associated with the weight of the cooled air draws air downward, warm air must be pulled into the gap by viscous stresses. For fluid entering a gap of thickness $h$, at speed $U$, the gradient of viscous stress can be estimated as: $\sim \eta U/h^2$, where $\eta$ is the viscosity of air. We estimate the velocity $U$ by balancing the viscous stress gradient with the buoyancy force; $\rho g \alpha \Delta T$ i.e.: $U \sim \rho g\alpha \Delta T h^2/\eta$. We then adopt the notation that if $f$ is a quantity of interest (e.g. temperature or gap height) that can vary over the pileus, then we write $f_r$ for the value of $f$ on the right edge of the pileus and $f_\ell$ for its value on the left edge of the pileus. If $U_r=U_\ell$ then there is the same inflow on the left and right edges of the pileus, and dispersal is symmetric and weak.

If there are different inflows on the left or right side of the pileus, then there can be net unidirectional flow beneath the pileus, carrying spores further. Assuming, without loss of generality, that the net dispersal of spores is rightward, the spreading velocity of the gravity current can be estimated from the difference: $U_g\sim U_\ell-U_r$ (right-moving inflow minus left-moving inflow). The furthest traveling spores originate near the rightward edge of the pileus and fall a distance $h_r$ (the gap width on the rightward edge) before reaching the ground. Since the gravity current spreads predominantly horizontally, the vertical trajectories of spores are the same as in still air, namely they sediment with velocity $v_s$ and take a time $\sim h_r/v_s$ to be deposited. By balancing the weight of a spore against its Stokes drag, we obtain $v_s = \frac{2\rho a^2g}{9\eta}$ where $\rho=1.2\times 10^3$ kg/m$^3$ is the density of the spore, and $a=2-4\,\mu$m is the radius of a sphere of equivalent volume\cite{Roper_10}. The sedimentation velocity, $v_s$, can vary between species, (typically $v_s=1-4\,$mm/s), but does not depend on the flow created by the pileus. The maximum spore dispersal distance is then:
\begin{equation}
d_{\rm max} =\frac{U_gh_r}{v_s}\sim \frac{g\alpha}{\eta} \left[\Delta Th^2\right]_r^\ell \frac{h_r}{v_s} \label{eq:scaling}
\end{equation}
where we use the notation $[f]^l_r$ to denote the difference in the quantity $f$ between the left and right sides of our model mushroom. Unidirectional dispersal therefore requires either that $\Delta T_\ell\neq \Delta T_r$ (i.e. there is a temperature gradient between the two sides of the pileus, Fig 4A middle panel) or $h_\ell \neq h_r$ (i.e. the mushroom is asymmetrically shaped, Fig 4A lower panel). These two cases can be distinguished by the dependence of dispersal distance upon the gap height: for a temperature-gradient induced asymmetry we predict: $d_{\rm max} \propto [\Delta T]^\ell_r h_r^3$ (and $U_g \propto [\Delta T]^\ell_r h_r^2$), whereas for a shape induced asymmetry: $d_{\rm max} \propto \Delta T [h]^\ell_r h_r^2$ (and $U_g \propto \Delta T [h]^\ell_r h_r$). Both scalings are validated by numerical simulations (Fig. 4B). When rescaled using Equation (\ref{eq:scaling}), data from simulations with different values of $h_\ell$, $h_r$, $[\Delta T]^l_r$, $[h]^l_r$ or $v_s$ all collapsed to two universal lines corresponding to temperature and height asymmetries (Fig. 4C,D). 

Experimental data from real mushrooms is fit well by $d_{\rm max}\sim h_r^2$ (Fig. 3). Additionally the scaling (\ref{eq:scaling}) accords with experimental observations that dispersal distance does not depend on pileus diameter or on the rate of spore release (Fig. S2). We can not directly confirm that there are not temperature gradients over the pileus surface. But these data are consistent with shape asymmetry; that is variation in the thickness of the pileus, and height of the gap, playing a dominant role in asymmetric spore dispersal from real mushrooms. Additionally, we made another direct test of our scaling law (\ref{eq:scaling}) by measuring $U_g$ (Fig. 4E, and see Materials and Methods). When the dimensionless prefactors in (\ref{eq:scaling}) were kept constant by using the same pileus, but the height of the gap, $h_r$, was varied, we found that $U_g\sim h_r$, consistent with the scaling for shape asymmetry-driven flow (Fig. 4F).

{\bf Spores can disperse over barriers surrounding the pileus}. In nature, pilei may grow crowded together or under plant litter or close to the host or substrate containing the parental mycelium (Fig 1A-B); thus, in addition to needing to disperse from the narrow gap beneath the pileus, spores may potentially also need to climb over barriers to reach external airflows. Although the cold, spore laden air is denser than the surrounding air, to replace the cold air that continuously flows from beneath the pileus, warm air must be drawn in to the pileus. Our simulations showed that when the gravity current met a solid barrier, the warm inflow and cold outflow could link to form a convective eddy (Figure S3). We studied whether this eddy could lift spores into the air, and what effect this might have upon spore dispersal. We found that when mushrooms were surrounded by a vertical barrier (see Materials and Methods), spores were dispersed over the barrier (Fig. 5A) provided that their horizontal range, predicted using Equation (1), exceeded the height of the barrier (Fig. 5B). 

Surprisingly, spores climbing over the barrier were dispersed apparently symmetrically in all directions from the top of the barrier (compare Fig. 5A with 1C) over the entire area of the box containing the mushroom and barrier (Fig. 5A). In particular, the total horizontal extent of the spore deposit (the size of the box) typically greatly exceeded the horizontal range predicted by Equation (1), even without accounting for the distance traveled by the gravity current in crossing over the barrier. 

The enhancement of spore dispersal by a barrier can be attributed to the action of the recirculating eddy: spores that climb the barrier may enter the current of air that is pulled down to the mushroom to replace the cold spore laden air (Fig. 5C). This eddy can carry spores up and further away from the barrier, and it is likely that no longer being constrained to travel along the ground surrounding the mushroom contributes to their increased range. 

Why does the height that gravity currents need to climb up the barrier not reduce their dispersal distance? In a climbing gravity current spores continue to sediment vertically downward. Since the vertical velocity of a climbing gravity current ($U_g\sim 4$ cm/s) is typically much larger than the sedimentation speed ($v_s\sim 1$ mm/s), spores do not sediment out of vertical gravity current as it climbs. To test this idea quantitatively, we numerically simulated spore dispersal by a gravity current that encounters a barrier inclined at an angle $\theta$ to the horizontal. Since the velocity of the gravity current is directed parallel to the barrier, spores can still sediment toward the barrier, but at a reduced sedimentation velocity $v_s\cos\theta$. (Sedimentation parallel to the barrier, with velocity $v_s\sin\theta$ can be neglected). Revisiting Equation (\ref{eq:scaling}) we therefore predict that spores will disperse a distance $U_g h_r/(v_s\cos\theta)$, i.e. they will travel a factor of $1/\cos\theta$ further up the slope than they would travel horizontally, and this prediction agrees quantitatively with numerical simulations (Fig. 5C). In particular, if $\theta=90^o$ (a vertical wall), we predict no sedimentation onto the wall, consistent with experimental observations that the distance that a gravity current climbs up a vertical wall does not reduce its horizontal spreading. 

Finally we note that, though our experiments were designed to exclude external winds, in nature wind speed tends to increase with height above the ground \cite{Aylor_90}, so spore that travel upward and then away from the vertical wall may be more likely to reach dispersive winds.

\section{Conclusion}

Mushrooms are not simple machines for producing the largest number of spores, but directly influence the dispersal of those spores, even during the phase of their dispersal previously considered to be passive. Rather than being the result of failure to evolve water-conserving adaptations, rapid water loss from the pileus enables mushrooms to create convective cells for dispersing spores. Convectively-generated winds provide a mechanism for spore dispersal for pilei that are crowded close together or close to the ground. Indeed, the presence of nearby boundaries for the upward flowing part of the current to climb may enhance spore dispersal.

Our analysis of convective dispersal adds to an increasing body of work revealing that fungal sporocarps are exquisitely engineered to maximize spore dispersal potential \cite{roper2008explosively,Fritz_13,Fischer_10_3}. Although evaporative cooling is an essential ingredient for spore dispersal and has been observed in many species \cite{Husher_99}, our analysis reveals the previously unreported role played by the asymmetric thickness of the pileus, or of the gap beneath it in shaping and amplifying the dispersive  wind created by the mushroom. Asymmetric spreading of spores was seen both in cultured and wild-collected mushrooms (respectively {\it L. edodes} and {\it A. californicus}) with circular pilei.

Although in some environments mushrooms may be able to rely on external winds to disperse their spores from the moment that spores are ejected, we have highlighted two common growing configurations in which there is likely to be little external wind in the gap beneath the pileus. There are an unknown number of such low-wind environments or niches for which the pileus generated wind may drive or assist spore dispersal. Moreover, the combination of controls that can be used by the fungus to manipulate dispersal distance (including gap width, asymmetry and temperature gradients) creates the potential for mechanistic explanations for why some species may disperse more effectively than others \cite{peay2012measuring,Norros_14}. 

\section{Materials and Methods}
\emph{Evaporation rate} We compared rates of water loss between {\it L. edodes} and {\it P. ostreatus} mushrooms and plants (one potted specimen of each of: {\it Tagetes erecta}, {\it Ocimum basilicum}, {\it Salvia elegans}, {\it Chrysanthemum} sp, {\it Dahlia variabilis}, acquired from Home Depot, cut leaves of {
\it Cerces canadensis}, {\it Pittosporum tobira}, {\it Tagetes erecta} and {\it Trachelospermum jasminoides}, collected on the UCLA campus). To measure the rate of evaporation from mushrooms with intact mycelia or from whole plants we wrapped the pot or the log on which the mushroom grew in plastic, so that water was lost only through the pileus / leaves. We also compared with water agar (1.5\% wt/vol) samples that were poured into petri dishes, and left uncovered. All samples were allowed to dry in a laboratory ambient (19$^\circ$C, 60\% R.H.). We measured surface area of samples by cutting the leaves and pilei into pieces and photographing the pieces.

\noindent \emph{Dispersal distance.} For spore dispersal measurements mushroom caps were placed in closed containers to isolate them from surrounding airflows. Depending on the gap height and length of gravity current, we used either 15cm diameter suspension dishes, food storage containers to which we added cardboard drop-ceilings, or file boxes. To prevent condensation from forming under the mushroom pileus we added a small quantity ($\sim$0.2\,g) of desiccant (DampCheck, Orlando, FL) to each container. Mushroom gap height was controlled by balancing the mushrooms on skewers spanning between two cardboard trestles, short lengths of tubing or wire connectors, or by suspending them from the ceiling of the container. We measured both the distance dispersed by the spores, and the asymmetry of the spore print. To calculate the print asymmetry in Figure 1D, we traced, using ImageJ, the coordinates of the boundary of the spore print. We then calculated asymmetry moments: $c_n = \int e^{ni\phi}\,dA$, where the integral is carried out over the area of the spore print, and $\phi$ is the angle made between the line joining each point to the centroid of the originating pileus and an arbitrary orientation axis. We characterized the asymmetry using a parameter ${\rm Asym} = \frac{|c_1|+|c_2|}{c_0}$. 

\noindent \emph{Velocimetry.} Spore dispersal was directly observed by illuminating the pileus with a laser sheet created by expanding a vertical laser beam from a 0.5W Hercules-450 laser (LaserGlow, Toronto, Canada) with a plano-concave lens (f=-7.7 mm, ThorLabs, Newton, N.J.). The mushroom pileus was placed inside a 18$\times$22$\times$22mm box: the ceiling and three walls of the box were lined with matte black aluminum foil (ThorLabs), and the fourth wall constructed of transparent acrylic, the floor of the container was covered with photography velvet to minimize scatter from the laser. The laser light sheet passed through a slit in the ceiling of the box to illuminate the rim of the mushroom cap. Spores traveling through the plane of the light sheet acted as bright tracer particles of the flow, and we filmed their movement using a FASTCAM SA3 high speed camera (Photron, San Diego, CA). Because of laser heating, after 2-4 s the laser created a thermal plume that obscured the airflows created by the mushroom. Accordingly, we analyzed only the first 2\,s of video captured from each experiment, and waited 5\,min to allow the surface to return to temperature equilibrium between runs of the experiment. Although our observation time was necesarily short, the time between observations was long enough for even slow overturning eddies to emerge. Spore velocities were measured by using a hybrid of particle imaging velocimetry (based on the code of \cite{sveen2004introduction}) to measure the velocities of groups of spores and individual particle tracking \cite{Roper:2013jr}.

\noindent \emph{Barrier crossing experiments} To measure the ability of mushroom-created airflows to disperse spores over barriers, we surrounded mushrooms by cylindrical walls, formed by taping a strip of transparency sheet inside of a 15cm diameter petri dish (Fig. 5A). By varying the width of the strip, we could vary the height of the barrier. To measure whether spores could cross this barrier, we placed transparency sheet on the floor of the petri dish (inside the barrier) and on the bench top outside of the barrier. To measure spore deposition just inside, and just outside of the barrier, we cut annuli of diameter 15mm from the two transparency sheets. The two annuli were cut into smaller pieces, and then separately vortexed in 10ml deionized water to wash off spores. The concentration of spores from inside and outside the barrier was counted automatically by the following method: Spore suspensions were added to a hemocytometer to create samples with known volume, which were photographed at 260$\times$ magnification using a Zeiss AxioZoom microscope. Spores appear as dark ellipses with bright outlines. Images were first denoised using median filtering and contrast enhanced using a Laplacian of Gaussian filter. Images were thresholded to identify bright edges and dark spore interior regions. Spores were segmented morphologically, by first removing all edges shorter than 30 pixels, and all dark regions with diameter smaller than 20 pixels or larger than 80 pixels. Spores were identified by dilating the dark regions using a disk element with a radius of 5 pixels, and keeping only those regions that overlapped with a detected bright edge. We then subtracted the bright edges off from the dark regions, to create one disconnected region per spore. The number of disconnected regions in each image was then counted. The ratio of the number of spores in the outer annulus to the number in the inner annulus gives a measure of the fraction of spores dispersed over the barrier. Since there is much larger total area of spore fall outside of the barrier than inside, the barrier crossing rate is proportional to, but not equal to, the fraction of spores that crossed the barrier.

\noindent \emph{Numerical simulations} To simulate the trajectories of spores in convective currents, we used the Boussinesq approximation. Specifically to calculate the inertia of the gravity current we take the density of air to be constant $\rho_0$, but when calculating the buoyancy force, we use a linear model: $\rho(T) = \rho_0 - \rho_0\alpha T$, where $T$ is the temperature measured relative to the ambient temperature far from the mushroom (so that $T\to 0$ far from the pileus), and $\alpha$ is a coefficient of expansion: $\alpha = 3.42\times 10^{-3}$/K. We non-dimensionalize our equations by scaling all distances by $L$, the half-width of the mushroom, all temperature differences by $\Delta T$, the maximum temperature drop at the mushroom surface, velocities by $U^* = (\alpha gL\Delta T)^{1/2}$ and pressure by $\rho_0U^{*2}$. The temperature and velocity fields around the pileus satisfy coupled partial differential equations (PDEs):
\begin{eqnarray}
\bu \cdot \nabla \bu &=& -\nabla p + \frac{1}{Re} \nabla^2 \bu +T \nonumber \\
\nabla\cdot\bu & =& 0 \label{eq:PDE}\\
\bu \cdot \nabla T &=& \frac{1}{P\acute{e}} \nabla^2 T \nonumber
\end{eqnarray}
 Spore trajectories were calculated by solving the ODEs: $\dot{x} = u$, $\dot{y} = v-v_s/U^*$, where $v_s$ is the spore sedimentation speed. Equations (\ref{eq:PDE}) contain two dimensionless numbers; the Reynolds number $Re\equiv \rho_0 U^*L/\eta$, formed from our velocity and length scales and the viscosity $\eta$ of air and the P\'{e}clet number $P\acute{e}\equiv U^*L/\kappa$ which depends on the thermal diffusivity, $\kappa$. In a typical simulation $Re=60$ and $P\acute{e}=40$. We used Comsol Multiphysics (COMSOL, Los Angeles) to set up and solve the PDEs in a 2D domain, in which the pileus was modeled by a semi-ellipse with a no-slip, constant temperature boundary condition on its surface. The pileus was set a small distance above a no-slip floor, and the domain was closed by a semi-circle on which the no-slip boundary condition was imposed, and whose diameter is 40$\times$ larger than the cap diameter. Constant temperature boundary conditions were applied on all surfaces, with $T=0$ on the walls of the domain, and cooling applied on the pileus surface (see main text for the two principal boundary conditions used). To avoid creating convective over-turning of fluid in the gap beneath the pileus, we isolated a section of the floor directly beneath the mushroom, and applied a zero-flux boundary condition there. To model the effect of nearby walls, we replaced the semi-circular external boundary by boundaries that were set a distance 1.5 from the mid-line of the mushroom, and oriented at an angle $\theta$ to the horizontal, as described in the main-text.

\begin{acknowledgments}
We thank Mechel Henry, Clive Roper, Christine Roper and Junius Santoso for experimental assistance and Mike Lawrence from UCLA Laser Safety for assistance with experimental design. L.Y. was supported by the Southern California Applied Math REU program (DMS-1045536). MR is supported by a fellowship from the Alfred P. Sloan Foundation, by NSF grant DMS-1312543 and by set-up funds from UCLA. E.D. is supported by set-up funds by NYU Polytechnic School of Engineering. We thank CCD Mushroom Inc. for providing the mushroom logs used in this experiment.
\end{acknowledgments}

\clearpage

\begin{figure}[H]
	\includegraphics[width=0.5\textwidth]{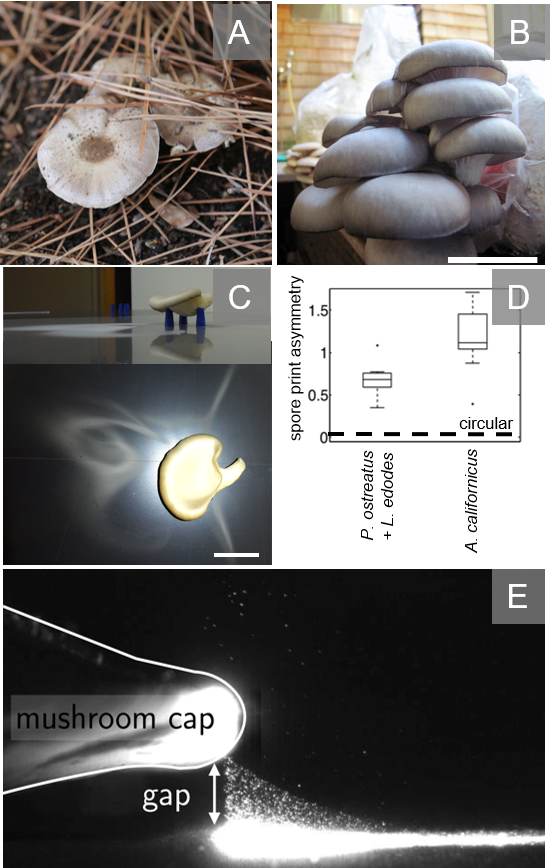}
\caption{Mushroom spores readily disperse from pilei that are crowded close together or close to the ground. (A) Oyster mushrooms ({\it Pleurotus ostreatus}) grow crowded together in a stand. (B) {\it Agaricus californicus} mushrooms grow under a layer of plant litter on the UCLA campus. (C) Even when mushrooms are isolated from external airflows, an asymmetric tongue of spores is deposited far from the pileus. Scale bar: 5cm. (D) Extended asymmetric spore deposits for cultured (left data) as well as wild-collected circular pilei (right data). The parameter used to characterize the asymmetry of the spore deposit is described in the Materials and Methods. (E) Airflows carrying spores out from the gap can be directly observed using a laser light sheet. Shown: {\it Lentinula edodes}, scale bar: 1 cm.}
\end{figure}

\begin{figure}[H]
	\includegraphics[width=0.5\textwidth]{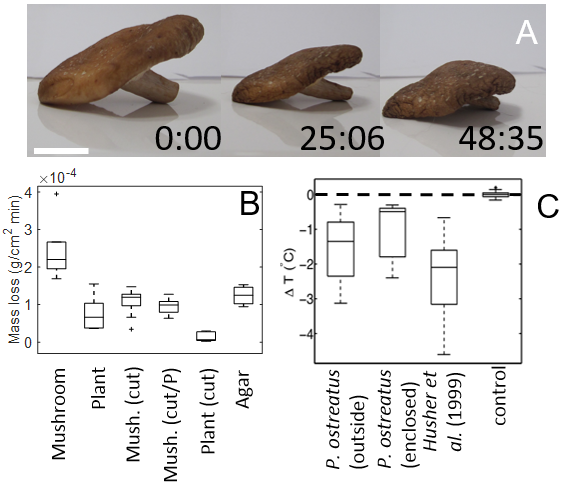}
\caption{High rates of evaporation from the pileus cool the surrounding air. (A) A {L. edodes} pileus left in laboratory ambient conditions rapidly dries out. Time code in {\it hh:mm} format. Scale bar: 1cm. (B) Rates of water loss from mushrooms greatly exceed plants. Mushrooms attached to intact mycelia (Mushrooms) lose water at a higher rate than living plants (Plants), and even agar hydrogels (Agar). Cut mushrooms (Mush. (cut)) lose water at a higher rate than cut plant leaves (Plant (cut)). Spore liberation contributed negligibly to mass loss: when the gill surface was coated with petroleum jelly to prevent spore shedding (Mush. (cut/P)), measured mass loss was statistically identical to untreated mushrooms. (C) Evaporation cools the air beneath the pileus by several degrees $^\circ$C, both for mushrooms stored in container to prevent external convection and for mushrooms maintained at laboratory ambient conditions, consistent with surface temperature measurements in \cite{Husher_99}. Also shown: a convection control showing temperature variations in one of our experimental containers with no mushroom.}
\end{figure}

\begin{figure}[H]
	\includegraphics[width=0.5\textwidth]{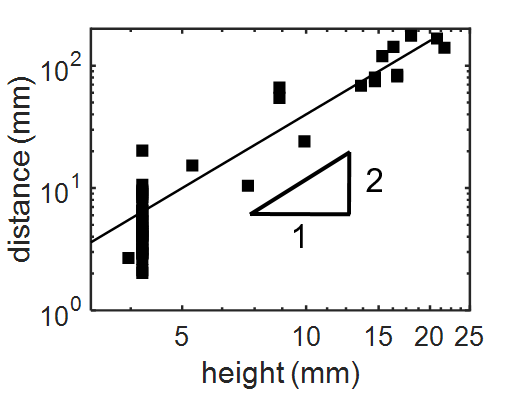}
\caption{Convective cooling produces a gravity current that disperses spores from beneath the pileus. Spore dispersal distance (black data points) increases as $($gap width$)^2$ (black line), consistent with theory for asymmetrically shaped pilei (Equation (1)).}
\end{figure}

\begin{figure}[H]
	\includegraphics[width=0.5\textwidth]{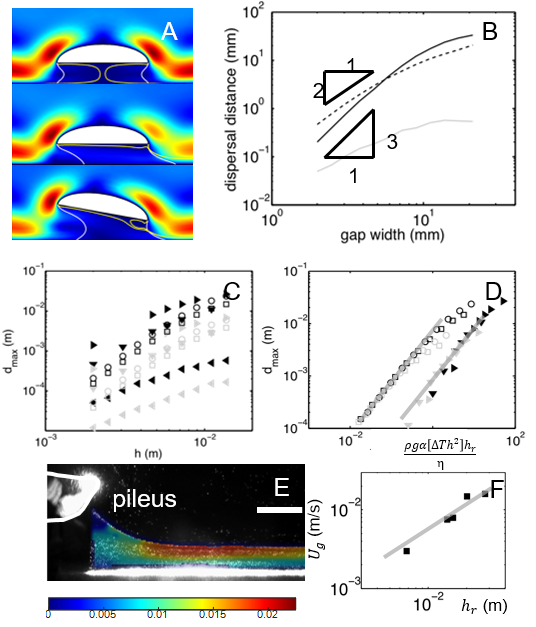}
\caption{Strong spore dispersal requires shape asymmetry or temperature differentials along the pileus. (A,B) Dispersal from a symmetrically cooled and shaped pileus is very weak (upper panel in (A), gray line in (B)). Imposing a temperature gradient across the pileus surface enhances dispersal (middle panel in (A), solid {black} line in (B)), as does asymmetric arrangement of the pileus (bottom panel in (A), dashed {black} line in (B)). Different forms of asymmetry produce different scalings for dispersal distance as a function of gap height. For temperature differentials, $d_{\rm max}\sim h_r^3$, while for shape asymmetry $d_{\rm max} \sim h_r^2$. Simulations are of a 4cm mushroom, with max surface $\Delta T=3^o$C, colors show speed of convective flow, white lines are representative trajectories of spores with sedimentation velocity 1mm/s, and yellow lines are flow streamlines. (C,D) A simple theory for the dispersal distance collapses data from different gap heights, spore sizes and amounts of asymmetry. (C) shows raw data for tilted pilei, with angles of tilt with the horizontal equal to 0$^o$ ($\blacktriangleleft$), 10$^o$ ($\blacktriangledown$), 20$^o$ ($\blacktriangleright$), and pilei with left-right temperature differentials of 1$^o$C ($\square$) and 1.5$^o$C ($\bigcirc$). Two sedimentation velocities are shown: $v_s=$ 1 mm/s (black symbols) and 4 mm/s (gray symbols). (D) shows same data collapsed using the scaling derived in the main text. Asymmetric pilei, and pilei with temperature differentials lie on different lines, but in both cases dispersal distance is proportional to $[h^2\Delta T]h_r/v_s$ (gray lines). (E,F) We test the scalings for real mushrooms using Digital PIV to measure spore velocities. In (E) colors give spore velocity in m/s, scale bar: 1cm). In (F), the mean spore velocity at the beginning of the gravity current is proportional to gap width $h$ (black line), consistent with Equation (1) for shape-induced asymmetry.}
\end{figure}

\begin{figure}[H]
	\includegraphics[width=0.45\textwidth]{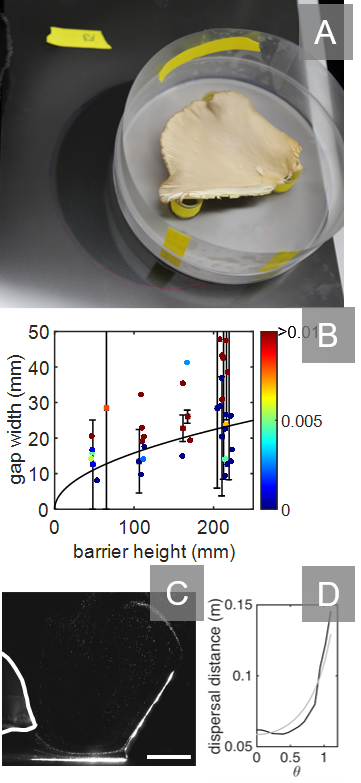}
\caption{Nearby boundaries enhance convective spore dispersal. (A) When a mushroom is surrounded by a circular barrier, spores can disperse over the barrier. Here we shifted the mushroom and barrier at the end of the experiment, so that the spore deposit can be seen more clearly by the contrast with the black circle of clean transparency beneath the barrier. In general spores that travel over the barrier were uniformly dispersed over the entire footprint of the box that enclosed the experiment, regardless of their horizontal spreading range. (B) Spores were able to cross a barrier if the height of the barrier was less than the horizontal range of spores (black line). Colors in the scatter plot correspond to a measure of barrier crossing rate (see Materials and Methods), in particular dark blue points correspond to mushrooms whose spores did not cross the barrier. (C) Spores climbing the barrier are observed being swept into the eddy of warm air that feeds the gravity current, potentially explaining their enhanced dispersal after crossing the barrier (shown: spore current from a {\it L. edodes} pileus). Scale bars: 2 cm. (D) Spores climbing an inclined barrier (black curve, numerical simulations) travel further than along horizontal surfaces (gray curve gives $1/\cos\theta\times$ the horizontal range of the spores).}
\end{figure}

\clearpage
\onecolumngrid

\begin{center}
\textbf{\large Supplementary Information for: ``Mushroom spore dispersal by convectively-driven winds''}
\end{center}
 \renewcommand{\thefigure}{S\arabic{figure}}
 \renewcommand{\thetable}{S\arabic{table}}
 \setcounter{figure}{0}
 
\begin{figure}[H]
	\begin{center}
	\includegraphics{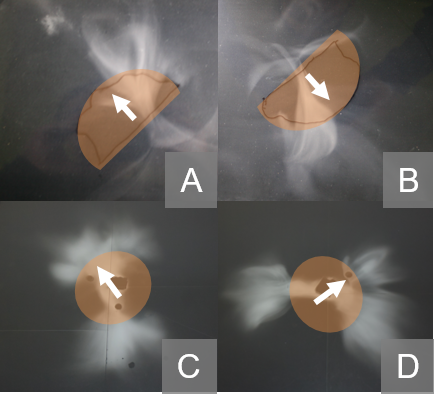}
	\end{center}
	\caption{To check that the airflows carrying spores out from under the pileus were created by the pileus, and not by external temperature gradients, we performed experiments in which the transparency was replaced in the middle of the experiment and the mushroom was rotated either by 90$^o$ or 180$^o$. In all cases we saw that the pattern of spore deposition rotated with the mushroom. Two representative experiments are shown here. The orange shapes show the approximate shape of the pileus, and arrows show orientation of arbitrary reference points on the pileus. (A-B) Deposition from a {\it P. ostreatus} mushroom. (A) Deposition after 2 hours. (B) The mushroom was rotated by 180$^o$, a new transparency added, and the experiment continued for another 2 hours. (C-D) Deposition from a {\it L. edodes} mushroom. (C) Deposition after 2 hours. (D) The mushroom was rotated by 90$^o$, a new transparency was added, and the experiment continued for another 2 hours.}
	\label{fig:convectioncontrol}
\end{figure}

\begin{figure}[H]
		\begin{center}
	\includegraphics{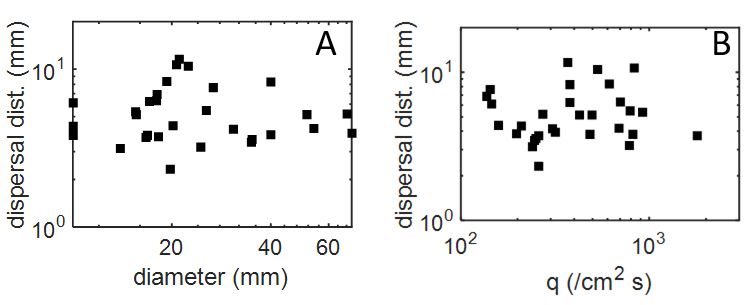}
		\end{center}
	\caption{Distance spores dispersed from under the pileus does not depend on the pileus diameter or on the rate of spore production. Here data from 30 {\it P. ostreatus} mushrooms are shown (a subset of the experiments from main text Fig. 3). (A) Dispersal distance does not correlate with pileus diameter ($R^2 =0.07$). (B) Dispersal distance does not correlate with the rate of spore production ($q$, number of spores released per cm$^2$ of pileus, per s, $R^2=0.17$).}
	\label{fig:dmaxversusdiameter}
\end{figure}

\begin{figure}[H]
		\begin{center}
	\includegraphics[width=4in]{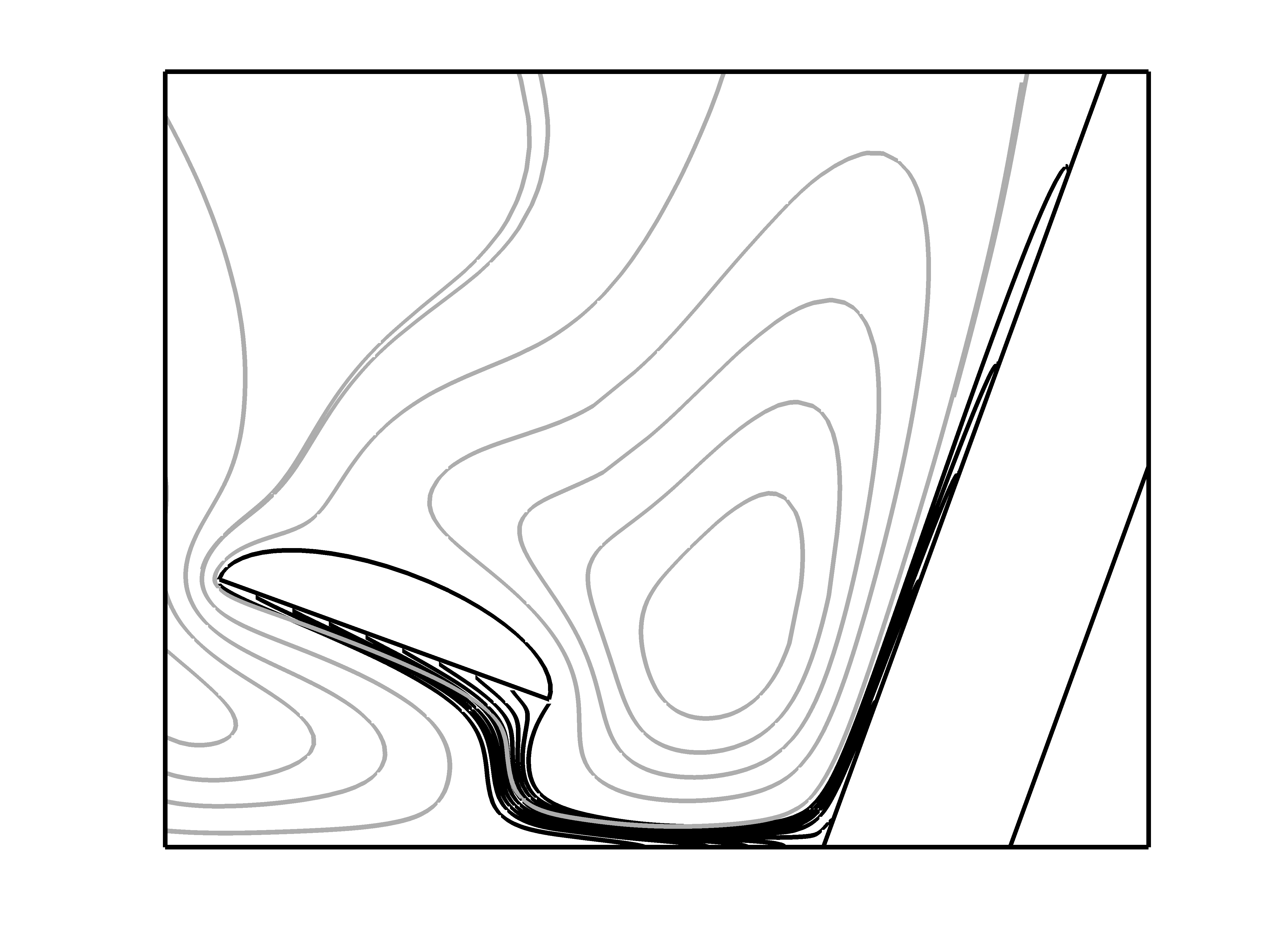}
		\end{center}
	\caption{Cold air leaving the gap beneath the pileus is replaced by pulling warm air to the pileus. In the presence of an external wall, numerical simulations (see main text: Materials and Methods), cold outflow and warm inflow can connect to form a closed convective eddy. Our simulations show that spores in the cold gravity current may be drawn up the wall by the closed eddy. Here, gray curves show the streamlines of the flow created by the cooled pileus. Black lines show simulated spore trajectories. The external wall is made up of two slanted surfaces on the right hand side of the image.}
	\label{fig:gravitycurrentmeetswall}
\end{figure}

 This document contains the following Supplementary Figures associated with the paper: ``Mushroom spore dispersal by convectively-driven winds'',
 
 \begin{itemize}
 	\item[\ref{fig:convectioncontrol}] Demonstration that spore deposition patterns rotate with the mushroom.
 	\item[\ref{fig:dmaxversusdiameter}] Spore dispersal distance does not depend on the pileus diameter or on the rate of spore production.
 	\item[\ref{fig:gravitycurrentmeetswall}] Numerical simulations show that gravity currents can carry spores up barriers.
 \end{itemize}

\begin{thebibliography}{10}
	
	\bibitem{kadowaki2010periodicity}
	Kadowaki K, Leschen RA, Beggs JR
	\newblock (2010) Periodicity of spore release from individual ganoderma
	fruiting bodies in a natural forest.
	\newblock \emph{Australasian Mycologist} 29:17--23.
	
	\bibitem{nagarajan1990long}
	Nagarajan S, Singh DV
	\newblock (1990) Long-distance dispersion of rust pathogens.
	\newblock \emph{Annual review of phytopathology} 28:139--153.
	
	\bibitem{Galante_11}
	Galante TE, Horton TR, Swaney DP
	\newblock (2011) $95\%$ of basidiospores fall within 1 m of the cap: a field-
	and modeling-based study.
	\newblock \emph{Mycologia} 103:1175.
	
	\bibitem{Roper_10}
	Roper M, {et~al.}
	\newblock ({2010}) {Dispersal of fungal spores on a cooperatively generated
		wind}.
	\newblock \emph{{Proceedings of the National Academy of Sciences}}
	{107}:{17474--17479}.
	
	\bibitem{roper2008explosively}
	Roper M, Pepper RE, Brenner MP, Pringle A
	\newblock (2008) Explosively launched spores of ascomycete fungi have
	drag-minimizing shapes.
	\newblock \emph{Proceedings of the National Academy of Sciences}
	105:20583--20588.
	
	\bibitem{Fritz_13}
	Fritz JA, Seminara A, Roper M, Pringle A, Brenner MP
	\newblock (2013) {A natural O-ring optimizes the dispersal of fungal spores}.
	\newblock \emph{J. R. Soc. Interface} 10:20130187--20130187.
	
	\bibitem{money1998more}
	Money NP
	\newblock (1998) More g's than the space shuttle: ballistospore discharge.
	\newblock \emph{Mycologia} pp 547--558.
	
	\bibitem{Pringle_05}
	Pringle A, Patek SN, Fischer M, Stolze J, Money NP
	\newblock (2005) The captured launch of a ballistospore.
	\newblock \emph{Mycologia} 97:866.
	
	\bibitem{Noblin_09}
	Noblin X, Yang S, Dumais J
	\newblock ({2009}) {Surface tension propulsion of fungal spores}.
	\newblock \emph{{Journal of Experimental Biology}} {212}:{2835--2843}.
	
	\bibitem{stolze2009adaptation}
	Stolze-Rybczynski JL, {et~al.}
	\newblock (2009) Adaptation of the spore discharge mechanism in the
	basidiomycota.
	\newblock \emph{PloS one} 4:4163.
	
	\bibitem{Buller_book}
	Buller AHR
	\newblock (1909) \emph{Researches on fungi}
	\newblock (London, New York [etc.] Longmans, Green and co.).
	
	\bibitem{Fischer_10_2}
	Fischer MWF, Stolze-Rybczynski JL, Cui Y, Money NP
	\newblock ({2010}) {How far and how fast can mushroom spores fly? Physical
		limits on ballistospore size and discharge distance in the Basidiomycota}.
	\newblock \emph{{Fungal Biology}} {114}:{669--675}.
	
	\bibitem{Norros_14}
	Norros V, {et~al.}
	\newblock (2014) {Do small spores disperse further than large spores?}
	\newblock \emph{http://dx.doi.org/10.1890/13-0877.1} 95:1612--1621.
	
	\bibitem{Aylor_90}
	Aylor DE
	\newblock (1990) {The role of intermittent wind in the dispersal of fungal
		pathogens}.
	\newblock \emph{Annual Review of Phytopathology} 28:73--92.
	
	\bibitem{Batchelorbook}
	Batchelor G
	\newblock (1967) \emph{Introduction to Fluid Dynamics}
	\newblock (Cambridge University Press, Cambridge, U.K.).
	
	\bibitem{Whitaker_10}
	Whitaker DL, Edwards J
	\newblock ({2010}) {Sphagnum Moss Disperses Spores with Vortex Rings}.
	\newblock \emph{{Science}} {329}:{406}.
	
	\bibitem{pedlosky1982geophysical}
	Pedlosky J
	\newblock (1982) \emph{Geophysical fluid dynamics}
	\newblock (Springer-Verlag, New York).
	
	\bibitem{mercier2014self}
	Mercier MJ, Ardekani AM, Allshouse MR, Doyle B, Peacock T
	\newblock (2014) Self-propulsion of immersed objects via natural convection.
	\newblock \emph{Physical Review Letters} 112:204501.
	
	\bibitem{Arora_86}
	Arora D
	\newblock (1986) \emph{{Mushrooms Demystified}}, A Comprehensive Guide to the
	Fleshy Fungi
	\newblock (Springer Science {\&} Business).
	
	\bibitem{Husher_99}
	Husher J, {et~al.}
	\newblock (1999) {Evaporative cooling of mushrooms}.
	\newblock \emph{Mycologia} 91:351--352.
	
	\bibitem{Mahajan_08}
	Mahajan PV, Oliveira F, Macedo I
	\newblock (2008) {Effect of temperature and humidity on the transpiration rate
		of the whole mushrooms}.
	\newblock \emph{Journal of Food Engineering} 84:281--288.
	
	\bibitem{baehr2011heat}
	Baehr H, Stephan K
	\newblock (2011) \emph{Heat and Mass Transfer}
	\newblock (Springer Berlin Heidelberg).
	
	\bibitem{Huppert_06}
	Huppert HE
	\newblock (2006) Gravity currents: a personal perspective.
	\newblock \emph{Journal of Fluid Mechanics} 554:299--322.
	
	\bibitem{Fischer_10_3}
	Fischer MWF, Money NP
	\newblock ({2010}) {Why mushrooms form gills: efficiency of the lamellate
		morphology}.
	\newblock \emph{{Fungal Biology}} {114}:{57--63}.
	
	\bibitem{peay2012measuring}
	Peay KG, Schubert MG, Nguyen NH, Bruns TD
	\newblock (2012) Measuring ectomycorrhizal fungal dispersal: macroecological
	patterns driven by microscopic propagules.
	\newblock \emph{Molecular Ecology} 21:4122--4136.
	
	\bibitem{sveen2004introduction}
	Sveen JK
	\newblock (2004) An introduction to matpiv v. 1.6. 1.
	\newblock \emph{Preprint series. Mechanics and Applied Mathematics http://urn.
		nb. no/URN: NBN: no-23418}.
	
	\bibitem{Roper:2013jr}
	Roper M, Simonin A, Hickey PC, Leeder A, Glass NL
	\newblock (2013) Nuclear dynamics in a fungal chimera.
	\newblock \emph{Proceedings of the National Academy of Sciences}
	110:12875--12880.
	
\end{thebibliography}
\end{document}